\documentclass[12pt]{iopart}

\usepackage{graphicx}
\usepackage{epsfig}

\begin{document}

\title{Is the assumption of a special system of reference consistent with
Special Relativity?}

\author{Rodrigo de Abreu and Vasco Guerra\footnote[1]{Also at Centro de F\'{\i}sica dos Plasmas, 
I. S. T., 1049-001 Lisboa, Portugal}
}
\address{Departamento de F\'{\i}sica, Instituto Superior T\'{ecnico}, 1049-001 Lisboa, Portugal}
\ead{vguerra@alfa.ist.utl.pt}


\begin{abstract}

In a previous work we have shown that the null result of the Michelson-Morley 
experiment in vacuum is deeply connected with the notion of time. The same is true for the 
the postulate of constancy of the two-way speed of light in vacuum in 
all frames independently of the state of motion of the emitting body. The argumentation 
formerly given
is very general and has to be true not only within Special Relativity and its `equivalence' of all 
inertial frames, but as well as in Lorentz-Poincar\'e scenario of a preferred reference frame. This 
paper is the second of a trilogy intending to revisit the foundations of Special Relativity, and addresses
the question of the constancy of the one-way speed of light and of the differences
and similarities between both scenarios. 
Although they manifestly differ in philosophy,
it is debated why and how 
the assumption of a ``special system of reference experimentally inaccessible" is indeed 
compatible with Einstein's Special Relativity, as beautifully outlined and discussed by 
John Bell \cite{Bell1988}. 
This rather trivial statement is still astonishing nowadays to a big majority of scientists.
The purpose of this work is to bring such assertion into perspective, widening the somewhat 
narrow view of Special Relativity often presented in textbooks.

\end{abstract}

\pacs{01.55; 03.30}



\maketitle

\section{Introduction}

In 2005 the whole world celebrated the 100th anniversary of Albert Einstein's annus mirabilis.
Of course the articles 
Einstein published in 1905 had a marked influence on Physics and Philosophy. But, 
more important, Einstein's thoughts and
his works keep feeding the scientific community, promoting debates, eventually some 
controversy, and compelling a permanent and comprehensive discussion of the foundations
of physics. Among the many events that took place during 2005, European Journal of Physics
published a focus section on ``Einstein and a Century of Time" \cite{EJP2005}, to which
we have also contributed \cite{GA2005}, confirming that 
Einstein's article ``On the electrodynamics of moving bodies" \cite{Einstein1905} keeps
inspiring innovative reflections on Special Relativity. 

In our previous contribution \cite{GA2005} we have shown that the roots of both
the null result of Michelson-Morley experiment and of the postulate of the constancy of the 
\textit{two-way} speed of light in vacuum
in all inertial frames, independently of the state of motion of the 
emitting body, rest firmly on the very notion of time.
In particular, we have explained that both can be obtained under 
three very reasonable assumptions: i) that all good clocks can be used to measure time, 
independently of the periodic physical phenomena
they are built upon; ii) that time is measured in the same way in all inertial frames, \textit{i.e.}, 
if a particular clock can be used to measure time in the ``rest system", a similar clock can be used to measure time in ``moving" inertial frames; iii) that a limit speed exists in the ``rest system". 
We noted in \cite{GA2005} that the arguments given are valid \textit{both} in the 
pre-relativistic Lorentz-Poincar\'e scenario of a preferred reference system and within Einstein's
relativity and its ``equivalence" of all inertial frames.
It is now time to address the question of a possible constancy of the \textit{one-way} speed of light,
which appears to be related to differences between these two scenarios.

Our reflection is essentially motivated by John Bell's analysis from \cite{Bell1988}. He has
noted the following (italics added in one sentence).
\begin{quote}
Many students never realize, it seems to me, that this primitive attitude, admitting a special system
of reference which is experimentally inaccessible, is consistent.

(...)
The approach of Einstein differs from that of Lorentz in two major ways. There is a difference
of philosophy, and a difference of style.

The difference of philosophy is this. Since it is experimentally impossible to say which of two
uniformly moving systems is \textit{really} at rest, Einstein declares the notions of
`really resting' and `really moving' as meaningless. For him only \textit{relative} motion
of two or more uniformly moving objects is real. Lorentz, on the other hand, preferred the view
that there is indeed a state of \textit{real} rest, defined by the `aether', even though the
laws of physics conspire to prevent us identifying it experimentally. \textit{The facts of physics
do not oblige us to accept one philosophy rather than the other.} 
And we need not accept
Lorentz's philosophy to accept a Lorentzian pedagogy. 
Its special merit is to drive home
lessons that the laws of physics in any \textit{one} reference frame account for all physical
phenomena, including the observations of moving observers. 
\end{quote}
From our experience, this is not a question of students only, as \textit{most scientists} never realize it 
either. It is striking the unease revealed by many colleagues when discussing the foundations
of Special Relativity and confronted with the compatibility between Einstein's results --
based on the notion of \textit{relative} motion -- and the
existence of a preferred reference frame -- with its associated idea of \textit{absolute} motion.

Due to the subject in question, it has to be clearly said this is not any kind of anti-relativistic paper.
In our journey into the foundations of Special Relativity, our aim is simply to use Lorentz's philosophy to explore the simple example of time dilation
and to consider the constancy of the one-way speed of light, in such a way that
no doubts can subsist on the formal compatibility of Lorentz and Einstein philosophies in
what concerns  the 
description of the physical  phenomena.
For this reason, the paper is written in a language corresponding to Lorentz's philosophy. 

The structure of the article
is as follows. In the next section the Lorentz-Poincar\'e view is established, together with 
its ``natural"
way of clock synchronization and the corresponding laws for transformation of coordinates
between inertial frames. Time dilation is illustrated in this framework.
Sections \ref{sec3} and \ref{sec4} introduce a formal Galileo transformation and the Lorentz
transformation, respectively. The description of time dilation is made again with the help of these 
transformations of coordinates. In section \ref{sec5} the question of the
constancy of the one-way speed of light is discussed, using all the 
previously presented coordinate transformations. As they describe one 
and the same reality, they are
obviously all mathematically equivalent, thus emerging the compatibility between
Einstein and Lorentz-Poincar\'e views in physical terms, although not in philosophical
ones. Finally, the last section summarizes the main conclusions of this work.

\section{External synchronization and the synchronized transformation}\label{sec2}

Within Lorentz-Poincar\'e view, the ``rest system" is the system in which
the one-way speed of  light in empty space is $c$ in any direction,
independently of the velocity of the source emitting the  light. This system is unique.
For the point we are trying to make here, it is irrelevant if this system is 
experimentally inaccessible or not, as the discussion
is essentially an academic discussion of principles, and not of direct practical interest.
Let us assume for now we do know which is the ``rest system". The consequence of 
not knowing it does not change the conclusions of this work and is examined in section \ref{sec6}.

The clocks from the ``rest system" can be synchronized with the usual Einstein procedure involving 
light rays \cite{Einstein1905}, since the one-way speed of light is known in this system.
Next, in order to perform time measurements in a ``moving" inertial frame (a frame
moving with constant speed in relation to the ``rest system"), 
it is necessary to
synchronize the ``moving" clocks. This cannot be done in the way used in the ``rest system", since in the ``moving" frame the one-way speed of light is not known (recall we are using the language 
of Lorentz's philosophy!). Nevertheless, it can 
easily be done with the help of the clocks ``at rest", because these clocks have already been synchronized.
Hence, the ``moving" clocks can 
be synchronized simply by ajusting them to zero whenever they fly past a clock ``at rest" that shows zero as well. From that moment on
the ``moving" clocks remain synchronous between themselves, thus establishing the common time of the ``moving system" (in order not to charge the notation, we will from now on drop the``" signs
in the words ``rest" and ``moving"). Evidently, this synchronization procedure is not the standard
one. We shall not start here any discussion around the ``conventionality of synchronization",
which has ample literature available and is only briefly addressed bellow.
A more comprehensive analysis of the subject is left for our final paper.

\begin{figure}
\begin{quote}
\begin{center}
\includegraphics[width=10cm]{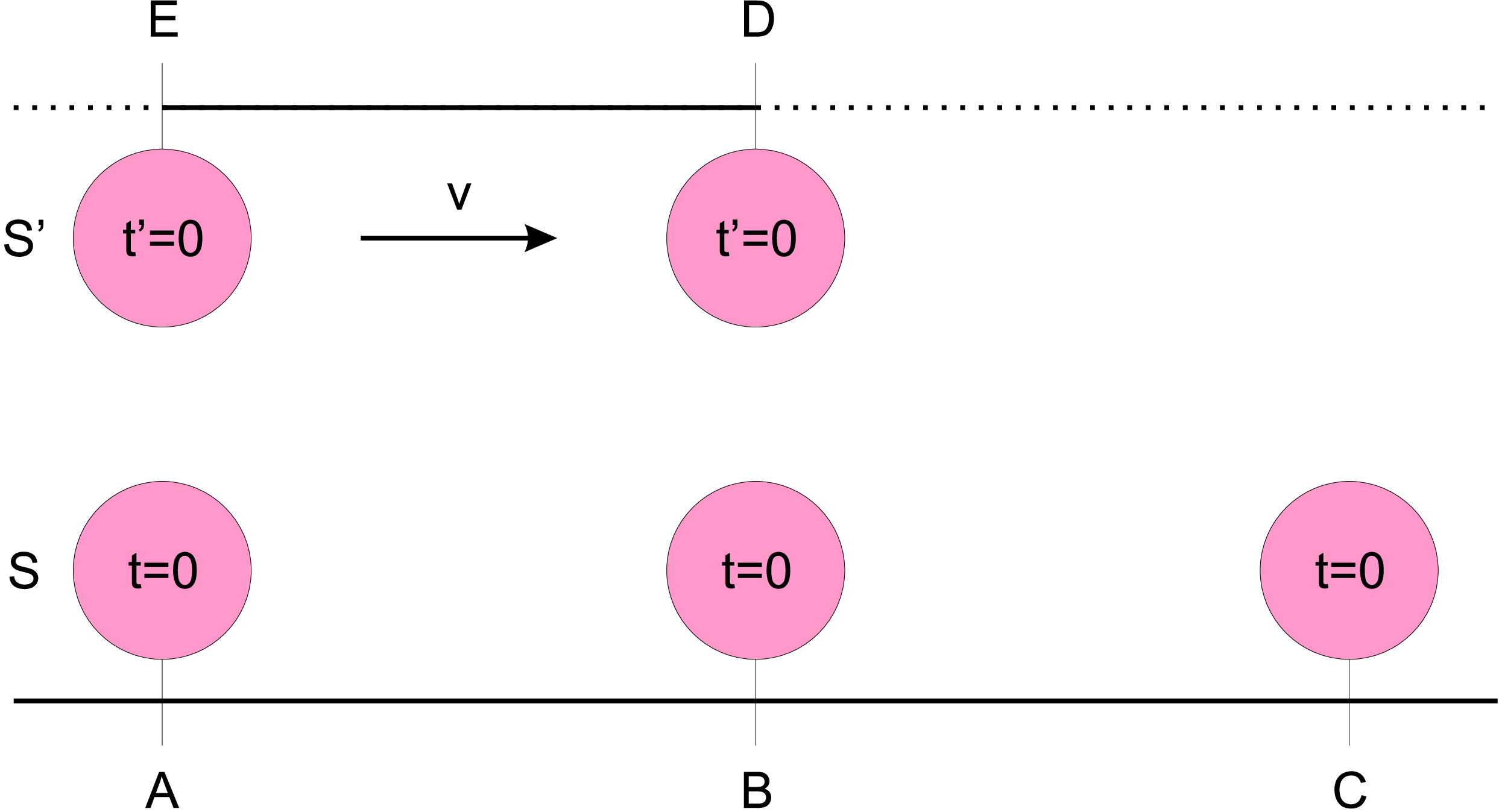}
\end{center}
\caption{Synchronization of clocks in a moving system: the moving clocks $D$ and $E$ are 
synchronized with the help of the previously synchronized clocks at rest $A$ and $B$.}\label{fig1}
\end{quote}
\end{figure}

Let $S$ denote the rest system and $S^\prime$ the moving one. For simplicity,
assume the axis of both frames are aligned, and that the origin of $S^\prime$ moves along the 
$x$-axis of $S$ with a certain speed $v$, in the
positive direction. The primed and non-primed quantities correspond to measurements made with the rulers 
and clocks of $S^\prime$ and $S$, respectively. The synchronization procedure just 
delineated corresponds simply to the statement that $t=0$ implies $t^\prime=0$, as it is shown in figure \ref{fig1}. Of course this synchronization method is an \textit{external} one, as noted
by Mansouri and Sexl in \cite{MS1977}, since to synchronize the clocks from $S^\prime$
one has to use the clocks from $S$.

\begin{figure}
\begin{quote}
\begin{center}
\includegraphics[width=10cm]{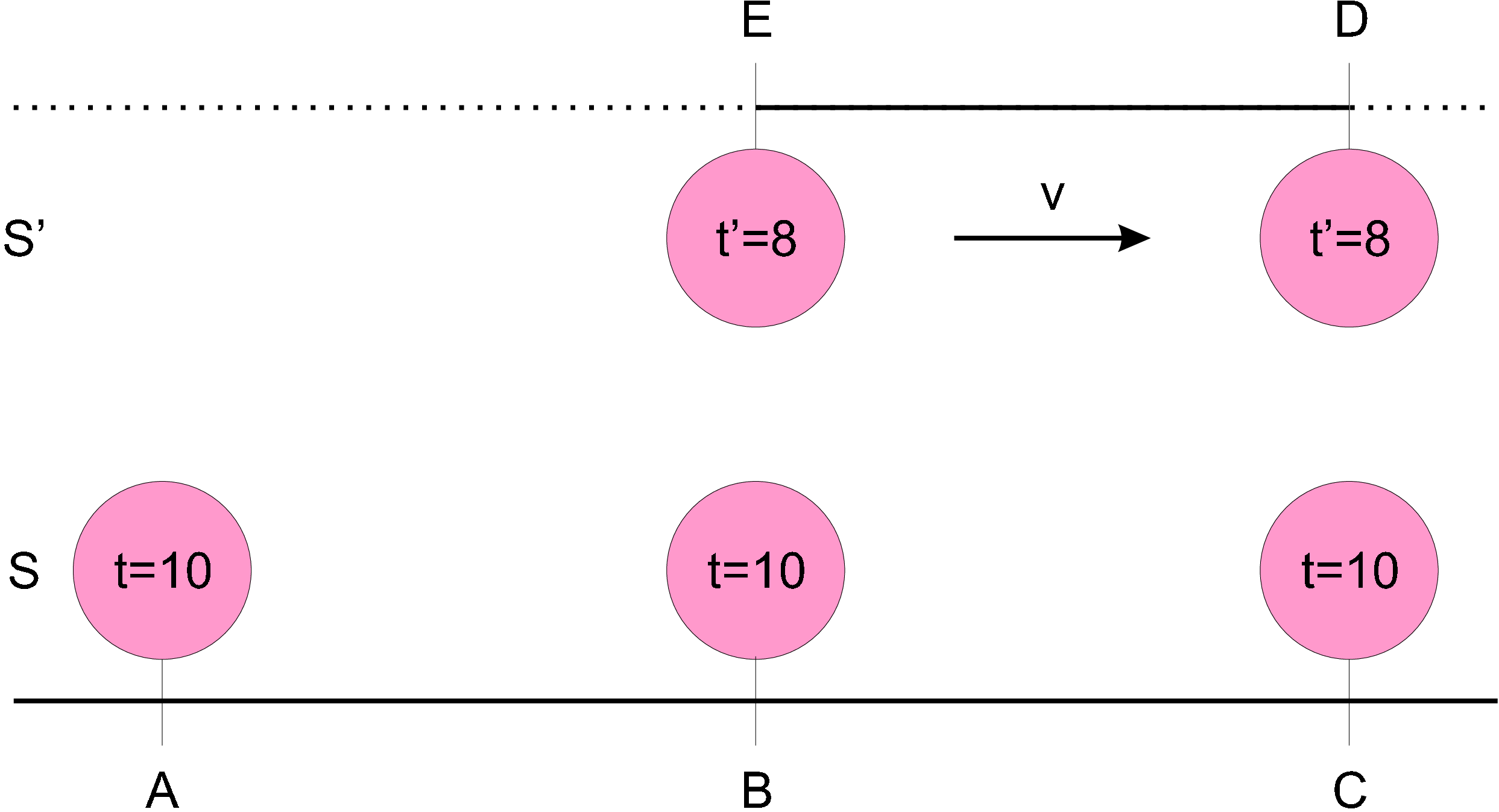}
\end{center}
\caption{Evolution of the situation from figure \ref{fig1} after 10 milliseconds. All times in the figure
are expressed in ms.}\label{fig2}
\end{quote}
\end{figure}

The phenomenon of time dilation can be deduced in the usual way, such as presented in
the classic textbooks from Feynman \cite{Feyetal1979} or Serway \cite{SB2000}, using
a light clock placed in $S^\prime$ aligned along the $y$-axis. The well-known result
\begin{equation}
\Delta t = \gamma t^\prime\ ,
\end{equation}
with
\begin{equation}\label{gamma}
\gamma=\frac{1}{\sqrt{1-\frac{v^2}{c^2}}}\ ,
\end{equation}
expresses the fact that ``moving clocks run slower". \textit{Each} of the moving clocks
experiences time dilation. Referring to figure
\ref{fig1}, if $v=0.6c$ (so that $\gamma=1.25$) and if the distances between clocks $A$ and $B$
and between clocks $B$ and $C$ are the same, $L$, and equal to 1800 km, the situation
depicted in figure \ref{fig1} evolves to the one depicted in figure \ref{fig2} at $t=10$ ms.
Always within Lorentz's philosophy, the effect is considered to be induced by absolute motion and clearly there is no reciprocity of time 
dilation. 

The phenomenon of space contraction can be deduced as well in the usual way, using
a light clock placed in $S^\prime$ aligned along the $x$-axis. The well-known result
\begin{equation}
L^\prime = \gamma L\ ,
\end{equation}
expresses the fact that ``moving rulers are shorter". Referring to figures
\ref{fig1} and \ref{fig2}, the distance in $S^\prime$ between clocks $D$ and $E$ is 
$L^\prime=\gamma L=1.25\times1800\ \mbox{km} = 2250$ km.
Again, the effect is considered to be induced by absolute motion and clearly there is no reciprocity of space contraction. 

The transformation of coordinates between $S$ and $S^\prime$ can now be readily obtained.
If the origins of both frames are considered to be at clocks $A$ and $E$, the position $x^\prime$
of clock $D$ in $S^\prime$ is simply given by
\begin{equation}
x^\prime = L^\prime = \gamma L = \gamma (x-vt)\ ,
\end{equation}
being $x$ its position in $S$. Consequently,
the relations between space and time coordinates
providing the translation from the description in the rest system to the one in a moving 
frame are just given by
\begin{eqnarray}
x^\prime & = & \gamma(x - v t) \nonumber \\
t^\prime & = & \frac{t}{\gamma} \label{TS}\ ,
\end{eqnarray}
where $\gamma$ is given by equation (\ref{gamma})
and $v$ is the absolute speed (\textit{i.e.}, the speed measured in the rest system) of 
the moving frame.

Expressions (\ref{TS}) form the \textit{synchronized transformation} and are very adequate
to analyses made within Lorentz's philosophy. They were obtained
by Mansouri and Sexl in 1977 \cite{MS1977} and have 
been emphasized by Franco Selleri since 1996 \cite{Selleri1996,Selleri2005},
who named them as \textit{inertial transformations}. 
Interestingly enough, the synchronized transformation is not symmetrical,
as the inverse transformation, expressing
$x^\prime$ and $t^\prime$ as functions of $x$ and $t$, is given by,
\begin{eqnarray}
x & = & \frac{1}{\gamma}(x^\prime + \gamma^2 v t^\prime) \nonumber \\
t & = & \gamma t^\prime\label{TSI}\ .
\end{eqnarray}
Notice that the position of the origin of $S$, $x=0$,
is given in $S^\prime$ by $x^\prime=-\gamma^2 v t^\prime$. 
This means that $S^\prime$ sees $S$ passing with speed $v^\prime=-\gamma^2 v$, and
not just $-v$ as one could think at first sight.
One factor $\gamma$ accounts for the fact that
rulers are shorter in $S^\prime$, while the second $\gamma$ factor comes from the fact that clocks run
slower there. 

It is not difficult to derive that 
if an object goes with absolute speed $w$, then its
relative speed, $w_v$, in relation to a frame $S^\prime$ moving with absolute speed $v$,
is given by
\begin{equation}\label{velad2}
w_v=\gamma^{\ 2}(w-v)=\frac{w-v}{1-\frac{v^2}{c^2}}\ .
\end{equation}
Thus, it is possible to calculate the one-way speed of light in a moving frame. If a light ray is emitted
and travels in the positive direction of the $x$-axis, we know it propagates in the rest system
with speed $c$,
independently of the speed of the source emitting the ray. In $S^\prime$, 
the one-way speed of this light ray is given by (\ref{velad2})
with $w=c$,
\begin{equation}\label{c+}
c_v^{\ +}=\gamma^{\ 2}(c-v)\ .
\end{equation}
If the light ray is emitted in the negative direction of the $x$-axis, $w=-c$ and its speed is given,
in absolute value, by
\begin{equation}
c_v^{\ -}=\gamma^{\ 2}(c+v)\ .
\end{equation}
Therefore, following Lorentz's philosophy, in the moving frame the one-way speed of light is not the same in different directions
and $c_v^{\ -}$ is always bigger than $c$. And yet the two-way speed of light is always $c$, as
the reader can easily verify.

Besides absolute time dilation and absolute space contraction, the 
synchronized transformation exhibits as well
absolute simultaneity. How can these statements be compatible with Special Relativity?
The answer is given in the next two sections.

\section{A formal Galileo transformation}\label{sec3}

Once the effects of time dilation and space contraction have been obtained, we can 
do something rather funny: since it is known that moving clocks run slower, it is easy to ``correct" the 
mechanisms of these clocks, so that they do not get delayed!
It is enough to tell the moving clocks $D$ and $E$ to mark always a factor $\gamma$
in advance of what they would mark without correction.

\begin{figure}
\begin{quote}
\begin{center}
\includegraphics[width=10cm]{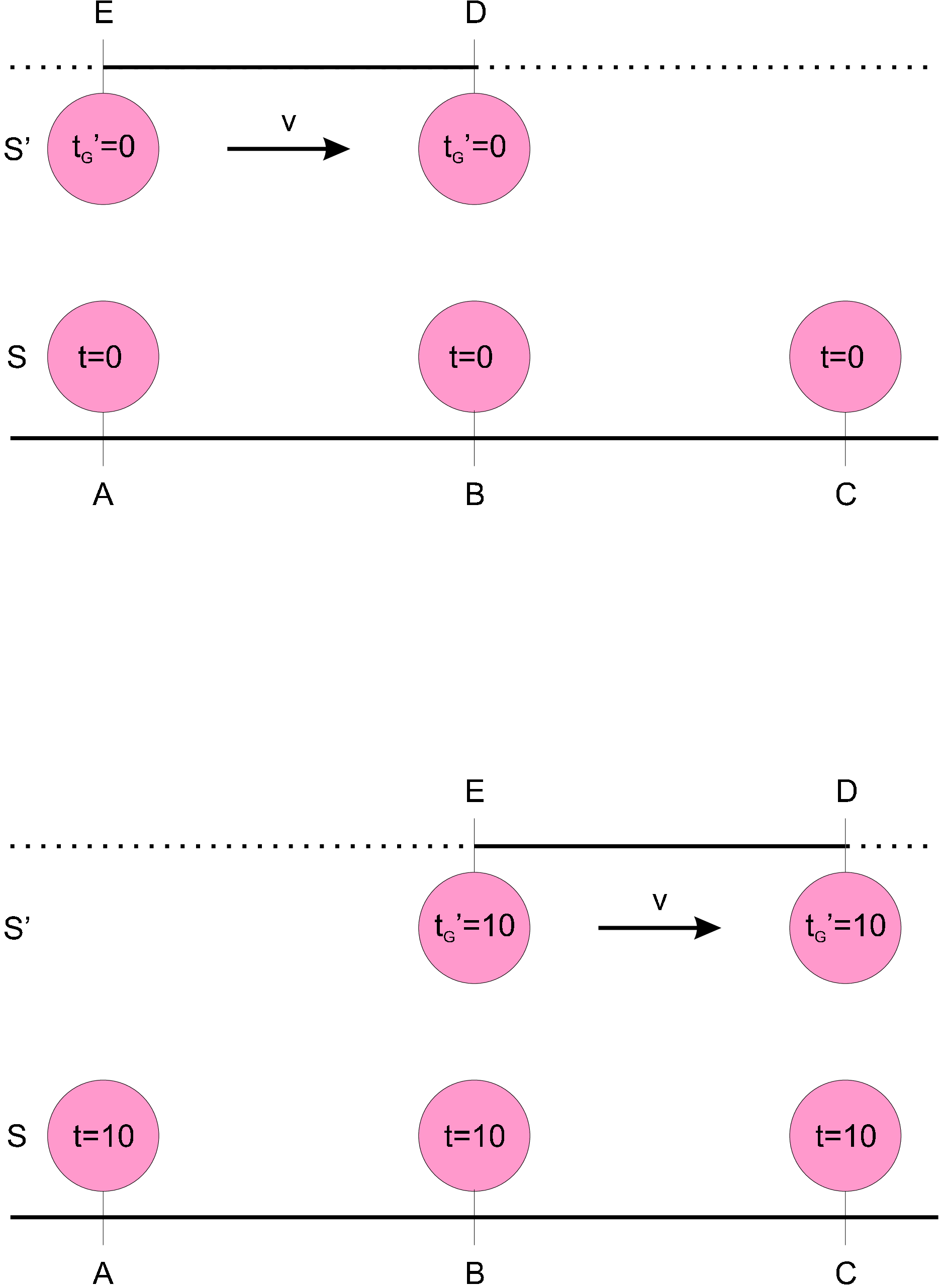}
\end{center}
\caption{``Time dilation" with synchronized Galilean clocks. All times expressed in ``miliseconds".}\label{fig3}
\end{quote}
\end{figure}

Suppose, for simplicity, that one ``tic-tac" of the clocks in $S$
corresponds  to ``1 second" (in $S$). Since in the present example
$\gamma=1.25$, one can simply \textit{define} one tic-tac from the clocks in $S^\prime$ as to be 1.25 seconds in
$S^\prime$. That being so, if $t_G^{\ \prime}$ denotes the ``time" given by these corrected clocks -- that we shall name 
\textit{Galilean clocks} -- then we simply have
\begin{equation}
t_G^{\ \prime}=t\ ,
\end{equation}
exactly as in the standard Galileo transformation. With Galilean clocks, the situation represented in figures \ref{fig1} and \ref{fig2}
is now depicted in figure \ref{fig3}. Galilean time \textit{appears} to be an absolute time, since it is the same in all 
inertial frames.

Of course figures \ref{fig1} and \ref{fig2} show precisely the \textit{same reality} as figure \ref{fig3}, 
which is simply \textit{described}
in a different way.
Nothing prevents the observers in $S^\prime$ from describing all events using Galilean clocks. There are a couple of problems, though.
The first one is that a clock working in a certain frame cannot be taken to another frame and be expected to work well. It will mark
wrong Galilean times. Before it can be used it must be corrected with the appropriate $\gamma$ factor, corresponding to the speed
of the new frame. Only then it can be utilized as a Galilean clock. The second question is related to the \textit{meaning} of the quantities measured with
Galilean clocks. Time itself, for a start. ``Time" given by Galilean clocks can be simply called ``time", but of course this does not
correspond to our intuitive notion of time and may originate some misunderstandings.

What has been done with clocks and time intervals can be done with rulers and lengths. We know that moving rulers are shorter. It is thus possible to correct the moving rulers so that they do not give shorter lengths. 
It is enough to tell the moving rulers to mark always a factor $\gamma$
less than what they would mark without correction.
If a working ruler is taken from $S$ to $S^\prime$, the mark for ``1 meter" should be replaced by $1/\gamma$ meters. That being so, if $L_G^{\ \prime}$ denotes the ``length" given by these 
corrected rulers -- that we shall name 
\textit{Galilean rulers} -- then we simply have 
\begin{equation}
L_G^{\ \prime}=L\ ,
\end{equation}
exactly as in the standard Galileo transformation.
Once more, nothing prevents the observers in $S^\prime$ from describing all events using Galilean rulers. This does not change the reality of what is being observed.

It is still possible to \textit{define} a \textit{Galileo speed}, $v_G$, as the ``speed" measured with Galilean clocks and Galilean rulers,
\begin{equation}\label{vG}
v_G=\frac{\Delta x_G}{\Delta t_G}\ .
\end{equation}
From the expressions given in section \ref{sec2}, it is easy to show that the ``Galileo speed" of an object whose absolute speed is
$w$, measured in an inertial frame $S^\prime$ going with absolute speed $v$, is just given by
\begin{equation}
v_G=w-v\ ,
\end{equation}
as in Galileo transformation. The ``Galileo speeds" of light are therefore 
\begin{equation}\label{cG+}
c_G^{\ +}=c-v
\end{equation}
and
\begin{equation}
c_G^{\ -}=c+v\ ,
\end{equation}
for light rays propagating in the direction of the movement of $S^\prime$ and in the opposing direction, respectively.
With Galilean clocks and Galilean rulers, the transformation of coordinates between $S$ and 
$S^\prime$ reduces 
formally to the Galileo
transformation,
\begin{eqnarray}
x_G^{\ \prime} & = & x - v t \nonumber \\
t_G^{\ \prime} & = & t\  \label{TGF}\ ,
\end{eqnarray}
where $v$ is the absolute speed of $S^\prime$.

It may be somehow surprising that this Galileo transformation is \textit{mathematically equivalent} to the synchronized transformation 
(\ref{TS}). Hence, in some sense it does not matter which of the transformations is used to \textit{describe}
the same reality. If we know the Galilean coordinates of
a certain event, then we can immediately know its synchronized coordinates, and vice-versa, as
long as the absolute speed of the moving frame is known.
The important question is not to assign an erroneous meaning to the measurements
that are being made. 

Two peculiarities of the Galileo transformation are still very interesting to note.
The first one is that, contrary to the synchronized transformation, Galileo transformation is \textit{symmetrical}.
As a matter of fact, the quantities in Einstein's frame can be expressed as a function of the Galilean ones by inverting (\ref{TGF}),
resulting
\begin{eqnarray}
x & = & x_G^{\ \prime} + v t_G^{\ \prime} \nonumber \\
t & = & t_G^{\ \prime}\  \label{TGFI}\ .
\end{eqnarray}
This set of equations is the same as (\ref{TGF}), simply interchanging the roles of the quantities in both frames and 
replacing $v$ by $-v$.
The position of the origin of $S$, $x=0$, is given in $S^\prime$ by $x_G^{\ \prime}=-v t_G^{\ \prime}$,
so that $S^\prime$ sees $S$ passing with \textit{Galileo speed} $-v$. In what concerns ``Galileo speeds", there is no
difference in the ways ``$S$ sees $S^\prime$" and ``$S^\prime$ sees $S$".
The second observation is
that if a second inertial frame $S^{\prime\prime}$ goes with absolute speed $w$, then the
transformation of Galilean coordinates between $S^\prime$ and $S^{\prime\prime}$ \textit{takes the same form} as between 
the rest system
and a moving inertial frame,
\begin{eqnarray}
x_G^{\ \prime} & = & x_G^{\ \prime\prime} + v_G t_G^{\ \prime\prime} \nonumber \\
t_G^{\ \prime} & = & t_G^{\ \prime\prime}\label{FGF1}\ ,
\end{eqnarray}
$v_G=w-v$ being the \textit{relative} Galileo velocity between both moving frames. 
These two facts could eventually suggest all moving inertial frames are equivalent to 
the rest system and only
relative motion is of importance. If
that would be the case,
the rest system would not be a privileged frame after all. But this ``equivalence" is purely formal. It 
emerges as a consequence of using Galilean clocks and rulers.

\section{The Lorentz transformation}\label{sec4}

According to Lorentz's philosophy, all the remarks made in the previous section for the
formal Galileo transformation can be made for the Lorentz transformation.
The latter provides only another way to relate the space and time coordinates of the rest system $S$
to the ones of a moving inertial frame $S^\prime$, but does not
have any fundamental privileged role. 
Once the clocks in both frames have been synchronized
as described in section \ref{sec2}, the Lorentz transformation can be easily obtained by ``correcting" in a particular way the time readings of the moving clocks. 

In the previous section Galileo transformation was attained by defining shorter 
seconds for the moving clocks. Now the rhythm of the clocks will not be changed, 
only the clocks will start from a 
different condition. Instead of adjusting
the moving clocks to mark $t^\prime=0$ when $t=0$ (see figure \ref{fig1}), we shall do that 
to one clock of $S^\prime$ only, which identifies the position $x^\prime=0$. The remaining moving 
clocks will be delayed by a factor that is proportional to their distance $x^\prime$ to the reference 
position $x^\prime=0$, which is given by
$\frac{v}{c^2}x^\prime$ (if $x^\prime$ is negative, this corresponds actually 
to advancing the clock). 
We shall denote the clocks altered in this way by \textit{Lorentzian clocks}, and their time readings, 
$t_L^{\ \prime}$,
by Lorentzian times. We thus have
\begin{equation}\label{Lt}
t_L^{\ \prime}=t^\prime-\frac{v}{c^2}x^\prime\ ,
\end{equation}
Why, following Lorentz's philosophy, should someone be interested in 
``de-synchronizing" clocks according to (\ref{Lt}) is related
to the problem of performing an \textit{internal} ``synchronization" of the moving clocks.
This will be fully clarified 
in the last paper of this trilogy. Anyway, as it is well known,
the resulting Lorentz transformation (see below)
is the natural transformation of coordinates used assuming an equivalence between
all inertial frames. Again we shall drop the ``" signs in the word ``synchronization", which
from now on denotes the external synchronization delineated in section \ref{sec2}.

\begin{figure}
\begin{quote}
\begin{center}
\includegraphics[width=10cm]{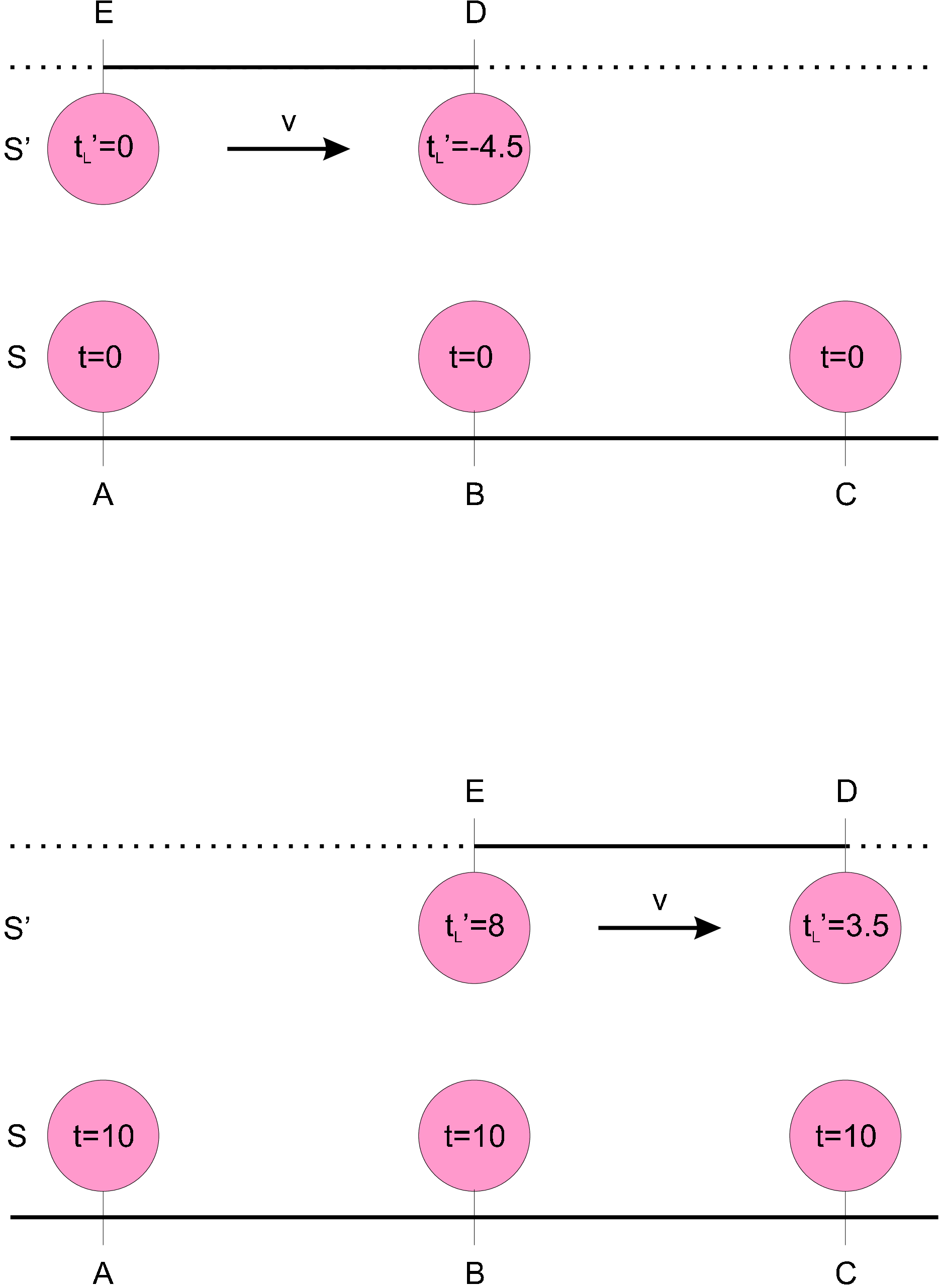}
\end{center}
\caption{Time dilation with de-synchronized and equal clocks. All times expressed in milliseconds.}\label{fig4}
\end{quote}
\end{figure}

Referring to figure \ref{fig1}, suppose clock $E$ defines the position $x^\prime=0$. Then, clock $E$ marks $t^\prime=0$
when $t=0$. Since clock $D$ is located at $x^\prime=2250$ km and $v=0.6 c$, it must be delayed
$(v/c^2)x^\prime=(0.6/c)\times2250\times10^3\simeq0.0045\ \mbox{s}=4.5$ ms. Therefore, at $t=0$ clock $D$ reads 
$t_L^{\ \prime}=-4.5$ ms.
Notice that the moving clocks $D$ and $E$ are exactly equal to clocks at rest $A$, $B$
and $C$, only they are not synchronized as to mark all $t=t_L^{\ \prime}=0$ at some arbitrary instant. The situation is shown in
the upper part of figure \ref{fig4}. Since the  moving clocks 
are precisely the same as in figures \ref{fig1} and \ref{fig2}, they exhibit
strictly the same time dilation as shown in those figures. Their \textit{rhythms} are affected 
by time dilation 
exactly as before. The only difference is that now their starting condition was set in a different way.
Hence, when 10 ms have passed in $S$, only 8 ms 
elapsed in $S^\prime$. 
More precisely, at $t=10$ ms clocks $D$ and $E$ mark $t_L^{\ \prime}=-4.5+8=3.5$ ms and 
$t_L^{\ \prime}=0+8=8$ ms, respectively, 
as represented in figure \ref{fig4}. 

With Lorentzian times, the expressions for transformation of coordinates between the rest
system and the moving frame
are easily found by substituting $t^\prime$ and $x^\prime$ given by the synchronized 
transformation (\ref{TS}) 
into (\ref{Lt}).
The \textit{Lorentz transformation} is finally obtained,
\begin{eqnarray}
x^\prime & = & \gamma(x-vt) \nonumber \\
t_L^{\ \prime} & = & \gamma\left(t-\frac{v}{c^2}x\right)\label{TL}\ ,
\end{eqnarray}
with $\gamma$ given by (\ref{gamma})
and $v$ denoting the absolute speed of $S^\prime$.

As with ``Galileo speed" (\ref{vG}), it is possible to \textit{define} an 
\textit{Einstein speed} (this speed could be named ``Lorentz speed" as well; however, we prefer to
call it Einstein speed, since it was defined and used by Einstein in his theory of relativity),
$v_E$, as the ``speed" measured with Lorentzian clocks (and ordinary rulers),
\begin{equation}\label{vE}
v_E=\frac{\Delta x}{\Delta t_L}\ .
\end{equation}
The time interval is calculated as the difference of
the ``time reading of a clock located at arrival position" with the ``time reading of a clock located 
at departure position".
Since Lorentzian clocks are de-synchronized, ``Einstein speeds" are
of course different from the ``true speeds" (which can be calculated by $v=\Delta x/\Delta t$ when the time intervals and distances 
are measured with synchronized clocks and ordinary rulers, respectively). 
Nevertheless, two-way speeds
of any object are the same with both types of clocks, since they are measured with one clock only and, that being so, any de-synchronization of distant clocks has no effect in the measurements.

Finally, the ``Einstein speed" $v_E$, measured in a frame moving with absolute speed $v$,
of an object which has absolute speed $w$, is  
\begin{equation}\label{vE2}
v_E=\frac{w_v}{1-\frac{vw_v}{c^2}} = \frac{w-v}{1-\frac{vw}{c^2}}\ .
\end{equation}
The ``Einstein speed" of light, $c_E$, exhibits a very interesting property. As a matter of fact, since the absolute
speed of light is always $c$, $c_E$ is obtained directly from (\ref{vE2}) with $w=c$:
\begin{equation}\label{cE}
c_E = \frac{c-v}{1-\frac{v}{c}} = c\ .
\end{equation}
Therefore, \textit{the ``Einstein speed" of light is always $c$ in
any moving inertial frame, independently of the speed of the moving frame}.

Notice that the remarks made about Galilean times, lengths and speeds can be repeated here.
Of course figures \ref{fig1} and \ref{fig2}, \ref{fig3}, and \ref{fig4}, show precisely the \textit{same reality}, which is simply \textit{described}
in a different way.
Even assuming Lorentz's philosophy,
nothing prevents the observers in $S^\prime$ from describing all events using Lorentzian clocks.
But this has to be done with care. For instance, two clocks placed in 
distinct locations working well in a certain frame cannot be ``transferred" 
to another frame and be expected to work well. They will mark wrong Lorentzian times. 
Before they can be used they must be corrected with the appropriate de-synchronization factors. Only then they can be utilized
as Lorentzian clocks.

It may be somewhat surprising that the Lorentz transformation 
(with its associated ``relativity of simultaneity", ``relativity of time dilation" and ``relativity
of space contraction"!)
is \textit{mathematically equivalent} to the 
synchronized transformation (\ref{TS}). That being so, \textit{any phenomenon described
by the Lorentz transformation can be described as well by the synchronized transformation}, which
is the natural transformation of coordinates in the Lorentz-Poincar\'e scenario of a preferred
frame. As a matter of fact,
if we know the Lorentzian coordinates of a certain event, then we can immediately know its synchronized 
coordinates, and vice-versa, as
long as the absolute speed of the moving frame is known. 
Reality is obviously independent of the choice of coordinates made to describe it.
The important question is not to assign an erroneous meaning to the measurements
that are being made.

As with the formal Galileo transformation, the Lorentz transformation is symmetrical (in what
concerns ``Einstein speeds", there is no difference in the ways ``$S$ sees $S^\prime$" and
``$S^\prime$ sees S") and the transformation of coordinates between two
moving inertial frames takes the same form as between the rest system and a moving
inertial frame (with $v_E$ in the place of $v$). But in Lorentz's philosophy this
equivalence is purely formal. It simply emerges as a consequence of using Lorentzian clocks.

\section{The one-way speed of light}\label{sec5}

The discussions about the ``constancy" of the one-way speed of light in all frames can 
now be easily understood.
Not only it is necessary to be careful with what is meant by ``constancy" (see \cite{GA2005}), but also
with what is meant by ``speed"! This should be
already clear and can be illustrated without effort with a simple example.

\begin{figure}
\begin{quote}
\begin{center}
\includegraphics[width=10cm]{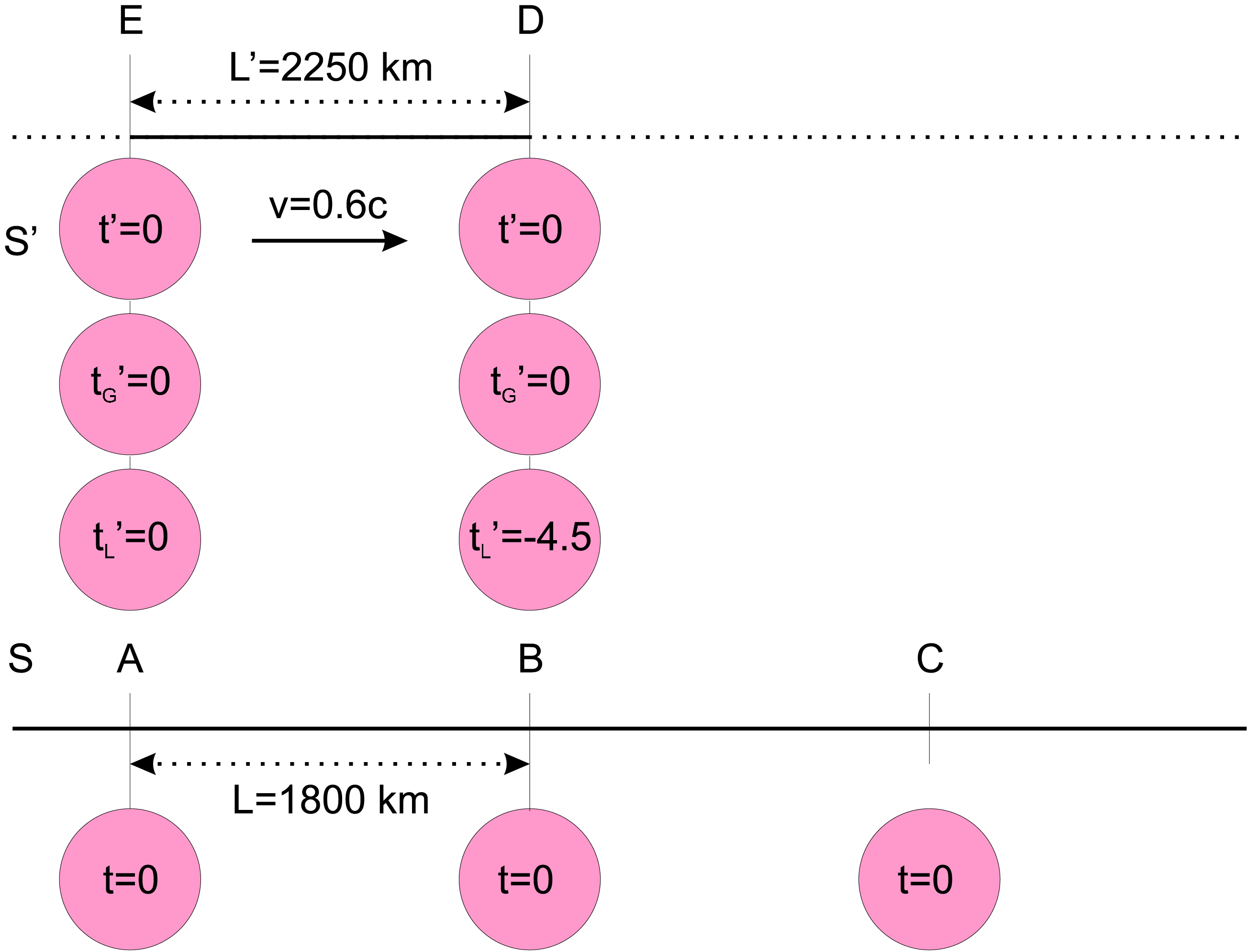}
\end{center}
\caption{Speed, ``Galileo speed" and ``Einstein speed". 
A light ray departs from point $E$ towards $D$ at $t=0$...}\label{fig5}
\end{quote}
\end{figure}

\begin{figure}
\begin{quote}
\begin{center}
\includegraphics[width=10cm]{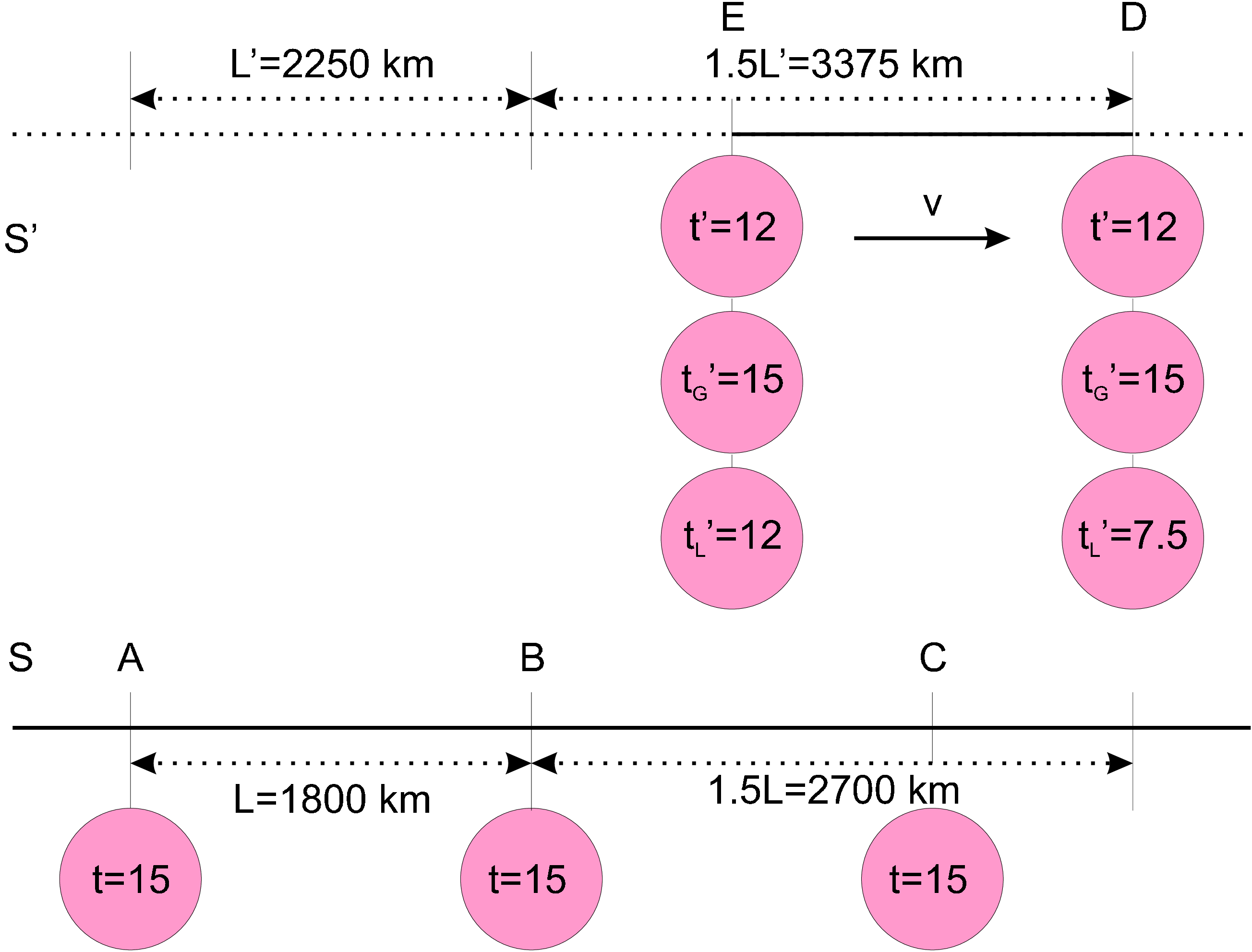}
\end{center}
\caption{... and arrives at point $D$ at $t=15$ ms. All times in the figure are expressed in milliseconds.}\label{fig6}
\end{quote}
\end{figure}

Suppose that the moving frame $S^\prime$ is equipped with all three types of clocks 
previously introduced (synchronized,
Galilean and Lorentzian), and that at $t=0$
a light ray is emitted from point $E$ to point $D$, as shown in figure \ref{fig5}. 
At what instant does the light ray reach $D$? In the ``rest system", $S$, light propagates with speed $c$, 
independently of the velocity of the source emitting the light. Since clock $D$ advances with speed $v$,
the relative speed (in $S$) of light and the clocks in $D$ is $c-v$, so that light takes $t=L/(c-v)\simeq
1800\times10^3/(0.4\times3\times10^8)=0.015$ s.
Therefore, when the light ray reaches $D$, the synchronized clocks in $S$
mark 15 ms. During this time interval, clocks at $D$ have advanced a distance 
$vt\simeq(0.6\times3\times10^8)\times(15\times10^{-3})=2.7\times10^6$ m, \textit{i.e.}, one and half times the distance
between clocks $B$ and $C$. It is still desirable to know which time readings the moving clocks show. The synchronized clocks
are affected by time dilation, hence marking $t^\prime=t/\gamma=15/1.25=12$ ms. Galilean clocks have been corrected
for the time dilation factor, and simply mark the same time as the rest clocks, $t_G=15$ ms. Finally, each of the Lorentzian
clocks has advanced the same 12 ms as the synchronized clocks. That being so, Lorentzian clocks $E$ and $D$ mark $0+12=12$ ms
and $-4.5+12=7.5$ ms, respectively. The situation corresponding to the arrival of the light ray at $D$
is represented in figure \ref{fig6}.
Now, what is left is to check the ``speeds" of light measured with the different sets of clocks,
according to the definition 
of speed as ratio between the ``length of the trip" with difference of
the ``time reading of a clock located at arrival position" with the ``time reading of a clock located 
at departure position". In $S$, the
speed of light is of course $c$,
\begin{equation}
v_S=\frac{(1800+2700) \mbox{km}}{(15-0)\mbox{ms}}=3\times10^8 \mbox{m/s} = c\ .
\end{equation}
In $S^\prime$ the speed of light should be given by equation (\ref{c+}), 
$\gamma_v^{\ 2}(c-v)=1.25^2\times(0.4c)=0.625c\simeq1.875\times10^8$ m/s. Do the synchronized
clocks from figures \ref{fig5} and \ref{fig6} provide this value? Indeed, since the distance
light travels in $S^\prime$ is just $L^\prime=2250$ km,
\begin{equation}
v_{S^\prime} = \frac{2250 \mbox{km}}{(12-0) \mbox{ms}} = 1.875\times10^8 \mbox{m/s}\ .
\end{equation}
With Galilean clocks and rulers, the ``Galileo speed" of light is given by (\ref{cG+}), $v_G=c-v=0.4c\simeq1.2\times10^8$ m/s.
Is that really so? Since Galilean rulers are corrected for space contraction, $L_G^{\ \prime}=L=1800$ km and
the ``Galileo speed" calculated from figures \ref{fig5} and \ref{fig6} is in fact
\begin{equation}
v_G=\frac{1800 \mbox{km}}{(15-0) \mbox{ms}} = 1.2\times10^8 \mbox{m/s}\ .
\end{equation}
Finally, ``Einstein speed" of light is obtained from the two figures as
\begin{equation}
v_E=\frac{2250 \mbox{km}}{(7.5-0) \mbox{ms}} = 3\times10^8 \mbox{m/s} \equiv c\ ,
\end{equation}
in accordance with the result (\ref{cE}). Notice that clock $D$ is showing only 7.5 ms at the arrival
of the light signal, but it was clock $E$ (and not $D$) that was marking $t_L^{\ \prime}=0$ at the departure of the signal.
Once more, it is meriting to emphasize that there is one and the same reality -- in this example
a light ray emitted from point $E$ 
to point $D$ -- that is simply being described in three different ways. This reality is of course independent
from the chosen description: they are all mathematically equivalent and it is easy to pass from one description to another.

What this example teaches us is the following: first, that the constancy of the one-way speed of light 
is a trivial statement regarding the ``Einstein speed" of  light; and second,
that such constancy is fully compatible with a non-constant one-way speed of light
as measured with the synchronized clocks. The Lorentz and Einstein philosophies correspond
to assigning the word ``time" to the time coordinates given by the synchronized transformation and
by the Lorentz transformation, respectively. The philosophies are different, but they are fully compatible
in physical terms.

\section{Summary and conclusion}\label{sec6}

The compatibility between Special Relativity and Lorentz-Poincar\'e view
of a preferred reference system experimentally inaccessible was shown to hold and
exemplified with the case of time dilation. The consistency of both scenarios 
is thoroughly discussed in \cite{AG2005}, where many other classic examples
are illustrated, such as the
``reciprocity" of time dilation and space contraction, the ``relativity of simultaneity",
the twin paradox, the problem of Bell's accelerating spaceships, the
propagation of spherical electromagnetic waves and the electric field of a moving point charge.

The key point is that
reality can be described in many different ways, for instance using ``synchronized", 
``Galilean" or ``Lorentzian" clocks, which are all mathematically equivalent. 
However, a change in the \textit{description} does not
change reality itself.
The question of the ``constancy" of the one-way ``speed" of light is essentially an issue
of \textit{language}, related to which 
description is being used. Therefore, when presenting or discussing Special Relativity we must keep in mind the precise meaning in which words like ``speed" and ``simultaneity" are used. 
Notice that this remark goes deeper than
the \textit{conventionalist} thesis and a vision of physics based on \textit{operationalism}.
Together with a discussion of the Principle of Relativity, these matters will be the
 subject of our third and final paper on the foundations of Special Relativity. 

The (eventual) impossibility of detecting experimentally the ``rest system" does not
change anything in Lorentz's philosophy. Within its framework, one would simply
have to admit he does not know the ``absolute speed" of an object, but this does not
promote the ``Einstein speed" to the status of ``true speed". They remain 
different notions. Consequently, the scenario of a preferred
frame keeps being consistent even if this frame cannot be identified, although its 
usefulness can be questioned. 
 
A very nice article with a message relatively similar to the one conveyed here
was published by Leubner and coworkers
in 1992 \cite{Leuetal1992}. After developing a non-standard synchronization, which they
named ``everyday synchronization" (and actually corresponds to the 
synchronized transformation for the particular case $v=-c$), they conclude:
\begin{quote}
 After these educational benefits of studying the set of standard `relativistic
 effects' also in everyday coordinates, we are of course happy to drop again
 `everyday' synchronization before proceeding to less elementary aspects of relativistic
 physics. For these aspects, we certainly prefer the more symmetric coordinate representations
 of expressions resulting from Einstein synchronization, but on purely
 practical grounds, and not on philosophical ones.
 \end{quote}
 
We subscribe both Bell's and Leubner's ideas, and hope in the near future
physics textbooks will regularly include an analysis of Lorentz's philosophy
and of its consistency, assuming without any prejudice 
that ``the facts of physics do not oblige us to accept one philosophy 
rather than the other". Regardless of the philosophy adopted, such approach
has various pedagogical advantages. Eventually, practical and even philosophical arguments 
can be invoked in favor of one philosophy. But, for strong and elegant
they may be, they remain practical and philosophical ones.

\end{document}